\documentclass[twocolumn,floatfix,preprintnumbers,nofootinbib,superscriptaddress]{revtex4}

\usepackage{ulem}
\usepackage{bm}
\usepackage{times}
\usepackage{amssymb,amsbsy,amsmath,amsfonts}
\usepackage{graphicx}
\usepackage{float}
\usepackage{color}
\usepackage{morefloats}
\usepackage{rotating}
\usepackage{srcltx}
\usepackage{slashed}
\usepackage{subfigure}
\usepackage{multirow}
\usepackage{verbatim}
\usepackage{hyperref}
\usepackage{tabularx}
\usepackage{braket}

\usepackage[outdir=./]{epstopdf}

\usepackage{graphicx}
\usepackage{amsmath}
\usepackage{amsfonts}
\usepackage{amssymb}
\usepackage{color}
\usepackage{multirow}
\usepackage{mathrsfs}
\usepackage{gensymb}

\usepackage{booktabs}  
\usepackage{threeparttable}

\usepackage{subfigure}

\newcommand\be{\begin{eqnarray}}
\newcommand\ee{\end{eqnarray}}

\begin{document}

\title{Understanding oscillating features of the time-like nucleon electromagnetic form factors within the extending vector meson dominance model}

\author{Bing Yan}
\affiliation{Institute of Modern Physics, Chinese Academy of Sciences, Lanzhou 730000, China}
\affiliation{College of Mathematics and Physics, Chengdu University of Technology, Chengdu 610059, China}

\author{Cheng Chen}
\affiliation{Institute of Modern Physics, Chinese Academy of Sciences, Lanzhou 730000, China}
\affiliation{School of Nuclear Sciences and Technology, University of Chinese Academy of Sciences, Beijing 101408, China}

\author{Xia Li}
\affiliation{College of Mathematics and Physics, Chengdu University of Technology, Chengdu 610059, China}

\author{Ju-Jun Xie}~\email{xiejujun@impcas.ac.cn}
\affiliation{Institute of Modern Physics, Chinese Academy of Sciences, Lanzhou 730000, China}
\affiliation{School of Nuclear Sciences and Technology, University of Chinese Academy of Sciences, Beijing 101408, China}
\affiliation{Southern Center for Nuclear-Science Theory (SCNT), Institute of Modern Physics, Chinese Academy of Sciences, Huizhou 516000, Guangdong Province, China}

\begin{abstract}

We investigate the nonmonotonic behavior observed in the time-like nucleon electromagnetic form factors. Using a phenomenological extending vector meson dominance model, where the ground states $\rho$ and $\omega$ and their excited states $\rho(2D)$, $\omega(3D)$, and $\omega(5S)$ are taken into account, we have successfully reproduced the cross sections of $e^+ e^- \to p \bar{p}$ and $e^+ e^- \to n \bar{n}$ reactions. Furthermore, we have derived the nucleon electromagnetic form factors in the time-like region, and it is found that the so-called periodic behaviour of the nucleon effective form factors is not confirmed. However, there are indeed nonmonotonic structures in the line shape of nucleon effective form factors, which can be naturally reproduced by considering the contributions from the low-lying excited vector states.

\end{abstract}

\maketitle

\section{Introduction}

Nucleons (proton and neutron) are composite particles consisting of three valence quarks ($uud$ and $udd$) and a neutral sea of strong interaction. Despite the discovery of the proton over 100 years, an exact theoretical description of its internal structure has not been achieved within the framework of quantum chromodynamics (QCD) theory. This is due to the non-perturbative nature of QCD in the energy regime of the nucleon. On the theoretical side, the nucleon electromagnetic structure can be described by the electromagnetic form factors (EMFFs), which depends on squared of the four-momentum ($q^2$) of the exchanged virtual photon. The EMFFs in the space-like region ($q^2 < 0$) are real and they can be associated with the charge and magnetic distribution of the nucleons~\cite{Pacetti:2014jai}. However, in the time-like region ($q^2 > 0$), the EMFFs become complex, providing information on the time evolution of the nucleon's charge and magnetic moments at the nucleon-antinucleon pair formation point~~\cite{Belushkin:2006qa,Kuraev:2011vq,Bianconi:2015owa,Tomasi-Gustafsson:2020vae}. On the experimental side, the EMFFs of nucleons were measured by both the $e^- N \to e^-N $ elastic scattering~\cite{JeffersonLabHallA:2001qqe,Gayou:2001qt,Andivahis:1994rq,Bosted:1992rq,JeffersonLabHallA:1999epl} and $e^+ e^- \to N \bar{N}$ annihilation reactions~\cite{BaBar:2013ves,Achasov:2014ncd,BESIII:2015axk,CMD-3:2015fvi,BESIII:2019tgo,BESIII:2019hdp,BESIII:2021rqk,BESIII:2021tbq,Huang:2021xte,SND:2022wdb}. One can obtain the space-like form factors from the former process and the time-like form factors from the latter process, respectively. 

Traditionally, most experimental data on electromagnetic form factors have been collected in the space-like region through electro-proton elastic scattering. However, a new era has begun with the introduction of electron-positron annihilation reactions~\cite{Ping:2013jka,BESIII:2017lkp,BESIII:2017ery}, where a nucleon-antinucleon pair is formed by a virtual photon.

Over the past decade, the BESIII Collaboration has made significant advancements in studying the timelike effective form factors of nucleons. Within the theoretical formula proposed in Refs.~\cite{Bianconi:2015owa,Tomasi-Gustafsson:2020vae}, it is found that not only the effective form factor of proton, but also the effective form factor of neutron has the periodic structures after subtracting a dipole or modified dipole contributions~\cite{BESIII:2021tbq}, leading to the discovery of an intriguing phenomenon known as oscillation behavior. This phenomenon has been further confirmed by recent measurements conducted by the SND Collaboration~\cite{SND:2022wdb}. However, these new measurements conflict with previously fitted results in the energy region below 2 GeV.

A comprehensive analysis of the nucleon's electromagnetic form factors in both the space- and time-like regions using dispersion theory is conducted in Ref.~\cite{Lin:2021xrc}. For the periodic behaviour of the nucleon's electromagnetic form factors, it has been studied in Refs.~\cite{Cao:2021asd,Xia:2021agf,Dai:2021yqr,Bianconi:2022yjq,Yang:2022qoy,Milstein:2022tfx,Tomasi-Gustafsson:2022tpu,PACETTI2022557,Qian:2022whn,Dai:2023vsw}, using various phenomenological methods. It is worthy mentioning that the periodic behaviour of the nucleon's electromagnetic form factors are induced by those broad vector mesons according to the studies in Ref.~\cite{Cao:2021asd}. However, these vector mesons couple to $N\bar{N}$ pair very weekly. In fact, a unified description of the time-like and space-like electromagnetic form factors of nucleons was proposed in Ref.~\cite{Lomon:2012pn}, where these low-lying vector mesons were taken into account. Meanwhile, in the time-like region, a revised Breit-Wigner formulas with momentum-dependent widths are needed for these vector resonances~\cite{Lomon:2012pn}. Indeed, the contributions of these vector mesons are important to the reactions of $e^+ e^-$ annihilation into light hadrons~\cite{BESIII:2019gjz,BESIII:2020kpr,BESIII:2021bjn,BESIII:2021yam,BESIII:2019ebn,BESIII:2020gnc,BESIII:2018ldc,BESIII:2020vtu,BESIII:2020xmw}.

The vector meson dominance (VMD) model is a successful approach for studying the baryon EMFFs, in both space-like and time-like regions~\cite{Iachello:1972nu,Iachello:2004aq,Bijker:2004yu,Yan:2023yff}. And, within the vector meson dominance model for studying the electromagnetic form factors of baryons, there is a phenomenological intrinsic form factor $g(s)$, which is a characteristic of valence quark structure. From these studies of the nucleon and hyperon EMFFs~\cite{Iachello:1972nu,Iachello:2004aq,Bijker:2004yu,Bianconi:2015vva,Yan:2023yff,Yang:2019mzq,Li:2021lvs}, it is found that a better choice of $g(s)$ is the dipole form
\begin{eqnarray}
g(s) = \frac{1}{(1-\gamma s)^2}, \label{eq:dipoleform}
\end{eqnarray}
with $\gamma=1.41$ $\rm GeV^{-2}$ for the case of nucleon, and $s$ is the invariant mass square of the $e^+ e^-$ annihilation process.

In this work, we study the $e^+e^- \to N\bar{N}$ reaction within the extended vector meson dominance model. Because proton and neutron are isospin doublets, the $p\bar{p}$ and $n\bar{n}$ states are expressed in terms of isospin 0 and 1 components. The mixtures of isovector and isoscalar for $p\bar{p}$ and $n\bar{n}$ of equal relative weight but different signs are imposed by the isospin symmetry as introduced by the underlying Clebsch-Gordan coefficients.In addition to these ground isovector $\rho$ meson and isoscalar $\omega$ meson, we will also study the important role played by the excited vector mesons with masses around 2.0 to 3.0 GeV,~\footnote{It's worth noting that due to the Okuba-Zweig-Iizuka rule~\cite{Okubo:1963fa,Zweig:570209,Iizuka:1966fk}, we do not consider the $\phi$ meson and its excited states here.} with the aim of describing the new experimental data reported by BESIII Collaboration.

Since the information about the $\rho$ and $\omega$ excited states around $2.0$ to $3.0$ GeV is scarce~\cite{ParticleDataGroup:2022pth}. Thus, it is necessary to rely on theoretical calculations. The BESIII Collaboration has also provided a large number of experimental data about $e^+e^-$ annihilation into light mesons~\cite{BESIII:2019gjz,BESIII:2020kpr,BESIII:2021bjn,BESIII:2021yam,BESIII:2019ebn,BESIII:2020gnc,BESIII:2018ldc,BESIII:2020vtu,BESIII:2020xmw}, which provides the basis for our construction of light vector states. From the analyses of the $e^+e^- \to \pi^+\pi^-$, $\omega \pi^0$, and $\rho^0 \eta'$ reactions, these $\rho$ excited sates were studied in Refs.~\cite{Wang:2020kte,Zhou:2022ark,Wang:2021gle}. The spectrum of excited $\rho$, $\omega$, and $\phi$ states were also investigated in Ref.~\cite{Wang:2021abg} using the modified Godfrey-Isgur model~\cite{Song:2015nia}. By performing the phenomenological analysis, the two-body strong decay of the excited $\rho$ and $\rho_3$ states are systematically studied~\cite{He:2013ttg}. Based on the above theoretical calculations and the vector mesons listed in the particle data group (PDG)~\cite{ParticleDataGroup:2022pth}, we take $\rho(2D)$, $\omega(3D)$, and $\omega(5S)$ into account in this work. Their masses and widths are collected in Table~\ref{tab:massandwidth}. By considering the contributions of these above low-lying vector meson excited states, we studied the $e^+e^- \to p \bar{p}$ and $e^+ e^- \to n \bar{n}$ reactions within the VMD model. It is found that the total cross sections of the two above reactions (corresponding to the nucleon effective form factors) and the nucleon electromagnetic form factors can be well described. However, the so-called oscillation behavior of the nucleon effective form factor discovered by the BESIII Collaboration~\cite{BESIII:2021tbq} is not confirmed.

\begin{table}[htbp]
    \centering
 \caption{Masses and widths of the excited vector states used in this work.}
    \begin{tabular}{c | c | c |c}\hline\hline
        State & Mass (MeV) & Width (MeV) & Reference \\\hline
        $\rho(2D)$ & $2040$ & $202$ & \cite{Zhou:2022ark,He:2013ttg}  \\
        $\omega(3D)$ & $2283$ & $94$ & \cite{Wang:2021gle} \\
        $\omega(5S)$ & $2422$ & $69$  & \cite{Wang:2021gle} \\
        \hline\hline
    \end{tabular}
       \label{tab:massandwidth}
\end{table}

It will be helpful to mention that the VMD model is a phenomenological approach and it works well in the low energy region. However, at high momentum transfer, the calculations from the VMD model are not consistent with that from the perturbative quantum chromodynamics~\cite{Friedman:1991nq,Brodsky:2016uln,Brodsky:2017icd}. In the present work, with the VMD model and the contributions from the low-lying excited $\omega$ and $\rho$ states, our calculations can give a reasonable description of the experimental measurements of the $e^+ e^- \to N\bar{N}$ reaction in the considered energy region. Meanwhile, our calculation offers some important clues for the reaction mechanisms of the process $e^+ e^- \to B \bar{B}$ and makes an effort to study the role of the vector states in relevant reactions.

This article is organized as follows: in the next section, we show the theoretical formalism of the nucleon EMFFs in the VMD model. Numerical results of the nucleon effective form factors, the total cross sections of $e^+e^- \to p\bar{p}$ and $e^+e^- \to n\bar{n}$ reactions, and the oscillating features of the nucleon EMFFs are shown in Sec.~\ref{sec:numerical results}. A short summary is given in the last section.

\section{Theoretical formalism} \label{sec:formalism}

Under the one photon exchange approximation, the total cross section of the reaction $e^+e^- \to N\bar{N}$ can be expressed in terms of the effective form factor $|G_{\rm eff}(s)|$ as~\cite{BaBar:2005pon,Dobbs:2014ifa,Haidenbauer:2014kja}
\begin{eqnarray}
 |G_{\rm eff}(s)| &=& \sqrt{\frac{2\tau|G_M(s)|^2+|G_E(s)|^2}{1+2\tau}}, \\
\sigma_{e^{+}e^{-}\rightarrow N \bar{N}} & =& \frac{4\pi\alpha^{2}\beta C_N}{3s}(1+\frac{1}
{2\tau})\mid G_{\rm eff} (s) \mid^2 ,
\end{eqnarray}
with $\alpha=e^2/(4\pi)$ the fine-structure constant, $\tau=q^2/(4m_N^2)$ and $\beta=\sqrt{1-4m_N^2/s}$ the phase-space factor. $m_N$ is the nucleon mass. Here, $C_N$ represents the S-wave Sommerfeld-Gamow factor accounts for the electromagnetic interaction of charged pointlike fermion pairs in the final state~\cite{Arbuzov:2011ff}. For proton, $C_p = y/(1-e^{-y})$ with $y=\frac{\alpha\pi}{\beta}\frac{2M_p}{\sqrt{s}}$, while for the neutron, $C_n = 1$. Considering the $C_N$ factor, it is expected that the total cross section of $e^+e^- \to p\bar{p}$ reaction is nonzero at the reaction threshold.

Based on parity conservation and Lorentz invariance, the electromagnetic form factors of the baryons with spin of $1/2$ can be characterized by two independent scalar functions $F_1(q^2)$ and $F_2(q^2)$ depending on $q^2$, which are called the Dirac and Pauli form factors, respectively. The electrical and magnetic form factors of nucleon can be written as~\cite{Irshad:2022zga,Sachs:1962zzc,Green:2014xba},
\begin{eqnarray}
    G_E(q^2)&=&F_1(q^2)+\tau F_2(q^2),\\
    G_M(q^2)&=&F_1(q^2)+F_2(q^2),
\end{eqnarray}
with $q^2 = s$. Once we have $F_1(s)$ and $F_2(s)$, we can naturally obtain the electromagnetic form factors of the nucleon. Then the total cross sections of $e^+ e^- \to N\bar{N}$ can be easily calculated.

Within the VMD model, the Dirac and Pauli form factors $F_1$ and $F_2$ for nucleon in the time-like region can be parameterized as,
\begin{eqnarray}
F_1^n &=& g(s) \left ( f_1^{n}-\frac{\beta_{\rho}}{\sqrt{2}}B_{\rho}-\frac{\beta_{\rho(2D)}}{\sqrt{2}}B_{\rho(2D)} +\frac{\beta_{\omega}}{\sqrt{2}}B_{\omega}  \right . \nonumber \\
&& \left .  + \frac{\beta_{\omega(3D)}}{\sqrt{2}}B_{\omega(3D)}+\frac{\beta_{\omega(5S)}}{\sqrt{2}}B_{\omega(5S)} \right ),   \\
F_2^n &=& g(s) \left ( f_2^{n}B_\rho-\frac{\alpha_{\rho(2D)}}{\sqrt{2}}B_{\rho(2D)} +\frac{\alpha_{\omega}}{\sqrt{2}}B_{\omega}  \right . \nonumber \\
    && \left . +\frac{\alpha_{\omega(3D)}}{\sqrt{2}}B_{\omega(3D)} +\frac{\alpha_{\omega(5S)}}{\sqrt{2}}B_{\omega(5S)} \right ), \\
F_1^p &=& g(s) \left ( f_1^{p}+\frac{\beta_{\rho}}{\sqrt{2}}B_{\rho}+\frac{\beta_{\rho(2D)}}{\sqrt{2}}B_{\rho(2D)} +\frac{\beta_{\omega}}{\sqrt{2}}B_{\omega} \right . \nonumber \\
    && \left . +\frac{\beta_{\omega(3D)}}{\sqrt{2}}B_{\omega(3D)}+\frac{\beta_{\omega(5S)}}{\sqrt{2}}B_{\omega(5S)} \right ),   \\
F_2^p &=& g(s) \left ( f_2^{p}B_\rho+\frac{\alpha_{\rho(2D)}}{\sqrt{2}}B_{\rho(2D)} +\frac{\alpha_{\omega}}{\sqrt{2}}B_{\omega} \right . \nonumber \\
    && \left . +\frac{\alpha_{\omega(3D)}}{\sqrt{2}}B_{\omega(3D)}  +\frac{\alpha_{\omega(5S)}}{\sqrt{2}}B_{\omega(5S)} \right ),
\end{eqnarray}
with
\begin{eqnarray}
	B_R &=& \frac{m_{R}^2}{m_{R}^2-s-im_{R}\Gamma_{R}}.
\end{eqnarray}
Here, $R \equiv \rho$, $\omega$, $\rho(2D)$, $\omega(3D)$, and $\omega(5S)$, and we take $m_{\rho} = 775.26$ MeV, $\Gamma_{\rho}=147.4$ MeV, $m_{\omega} = 782.66$ MeV, and $\Gamma_{\omega}=8.68$ MeV.

In addition, at $q^2=0$, with the constraints $G_E^{n}=0$ and $G_M^{n}=\mu_n$, $G_E^{p}=1$ and $G_M^{p}=\mu_p$, the coefficients $f_1^{n}$, $f_2^{n}$, $f_1^{p}$ and $f_2^{p}$ can be calculated, 
\begin{eqnarray}
f_1^n &= & \frac{\beta_{\rho} - \beta_{\omega}}{\sqrt{2}} + \frac{\beta_{\rho(2D)} - \beta_{\omega(3D)} - \beta_{\omega(5S)}}{\sqrt{2}},  \\ 
f_2^n &=& \mu_n -\frac{\alpha_{\omega}}{\sqrt{2}} + \frac{\alpha_{\rho(2D)} - \alpha_{\omega(3D)} - \alpha_{\omega(5S)}}{\sqrt{2}}, \\
f_1^p &=& 1- \frac{\beta_{\rho} + \beta_{\omega}}{\sqrt{2}} - \frac{\beta_{\rho(2D)} + \beta_{\omega(3D)} + \beta_{\omega(5S)}}{\sqrt{2}},  \\
f_2^p &=& \mu_p- 1 -\frac{\alpha_{\omega}}{\sqrt{2}} -\frac{\alpha_{\rho(2D)} + \alpha_{\omega(3D)} + \alpha_{\omega(5S)}}{\sqrt{2}},
\end{eqnarray}
with $\mu_n=-1.91$ and $\mu_p=2.79$ as quoted in the PDG~\cite{ParticleDataGroup:2022pth}. These coefficients $\beta_{\rho}$, $\beta_{\omega}$, $\beta_{\rho(2D)}$, $\beta_{\omega(3D)}$, $\beta_{\omega(5S)}$, $\alpha_{\omega}$, $\alpha_{\rho(2D)}$, $\alpha_{\omega(3D)}$, $\alpha_{\omega(5S)}$ are model parameters, which will be determined by fitting them to the experimental data of the $e^{+}e^-\to p\bar{p}$ and $e^+ e^- \to n\bar{n}$ reactions.

In the spacelike region, taking $Q^2 = -q^2 >0$, with the dipole form for $g(Q^2)$, it is found that for the large value of $Q^2$, $F_1 \sim \frac{1}{Q^4}$, and $F_2 \sim \frac{1}{Q^6}$, which are consistent with the asymptotic behavior of $F_1$ and $F_2$ predicted by the perturbative quantum chromodynamics~\cite{Brodsky:1973kr,Lepage:1979za,Lepage:1980fj}.

\section{NUMERICAL RESULTS} \label{sec:numerical results}

Under the above formulations, we perform the $\chi^2$ fit to the experimental data of $e^+ e^- \to N\bar{N}$ reaction. There are 224 data points in total, with center-of-mass energies range from the kinematic reaction threshold to 3.2 GeV. These experimental data are: i) the total cross sections of $e^{+}e^-\to N\bar{N}$ reaction from BESIII Collaboration~\cite{BESIII:2015axk,BESIII:2019tgo,BESIII:2019hdp,BESIII:2021rqk,BESIII:2021tbq}, CMD3 Collaboration~\cite{CMD-3:2015fvi}, $BABAR$ Collaboration~\cite{BaBar:2013ves}, and SND Collaboration~\cite{Achasov:2014ncd,SND:2022wdb,SND:2023fos}; ii) the electromagnetic form factors $|G_E|$ and $|G_M|$ for nucleon from BESIII Collaboration~\cite{BESIII:2019hdp,BESIII:2022rrg}; iii) the ratio of $|G_E/G_M|$ from $BABAR$ Collaboration~\cite{BaBar:2013ves}, CMD3 Collaboration~\cite{CMD-3:2015fvi}, and BESIII Collaboration~\cite{BESIII:2019tgo,BESIII:2021rqk}.

\begin{table}[htbp]
    \centering
 \caption{Values of the fitted parameters.}
    \begin{tabular}{c | c | c |c}\hline\hline
        Parameter & Value & Parameter & Value \\\hline
        $\beta_{\rho}$ & $1.471 \pm 0.131$ & $\alpha_{\omega}$ & $-904.176 \pm 20.211$  \\
        $\beta_{\omega}$ & $7.357 \pm 0.102$ & $\alpha_{\rho(2D)}$ & $0.601 \pm 0.101$ \\
        $\beta_{\rho(2D)}$ & $-0.419 \pm 0.109$ & $\alpha_{\omega(3D)}$ & $-0.734 \pm 0.081$ \\
        $\beta_{\omega(3D)}$ & $0.726 \pm 0.083$  & $\alpha_{\omega(5S)}$ & $0.313 \pm 0.028$ \\
        $\beta_{\omega(5S)}$ & $-0.334 \pm 0.034$ &  &  \\\hline\hline
    \end{tabular}
       \label{tab:fittedparameters}
\end{table}

The results of the fitted parameters are listed in Table~\ref{tab:fittedparameters}. The corresponding $\chi^2/\rm d.o.f$ is 1.6, where ${\rm d.o.f}$ is the number of dimension of the freedom. Note that the inclusion of more vector states will improve the fitting results, since it has more freedoms. We have explored such a
possibility, but we have found some fitting problems. The fitted parameters have large uncertainties and strong correlations, which means that the inclusion of more vector states would not significantly improve the fitted results. Thus, we consider only the $\rho(2D)$, $\omega(3D)$, and $\omega(5S)$ states here.

\begin{figure}[htbp]
	\centering
	\includegraphics[scale=0.4]{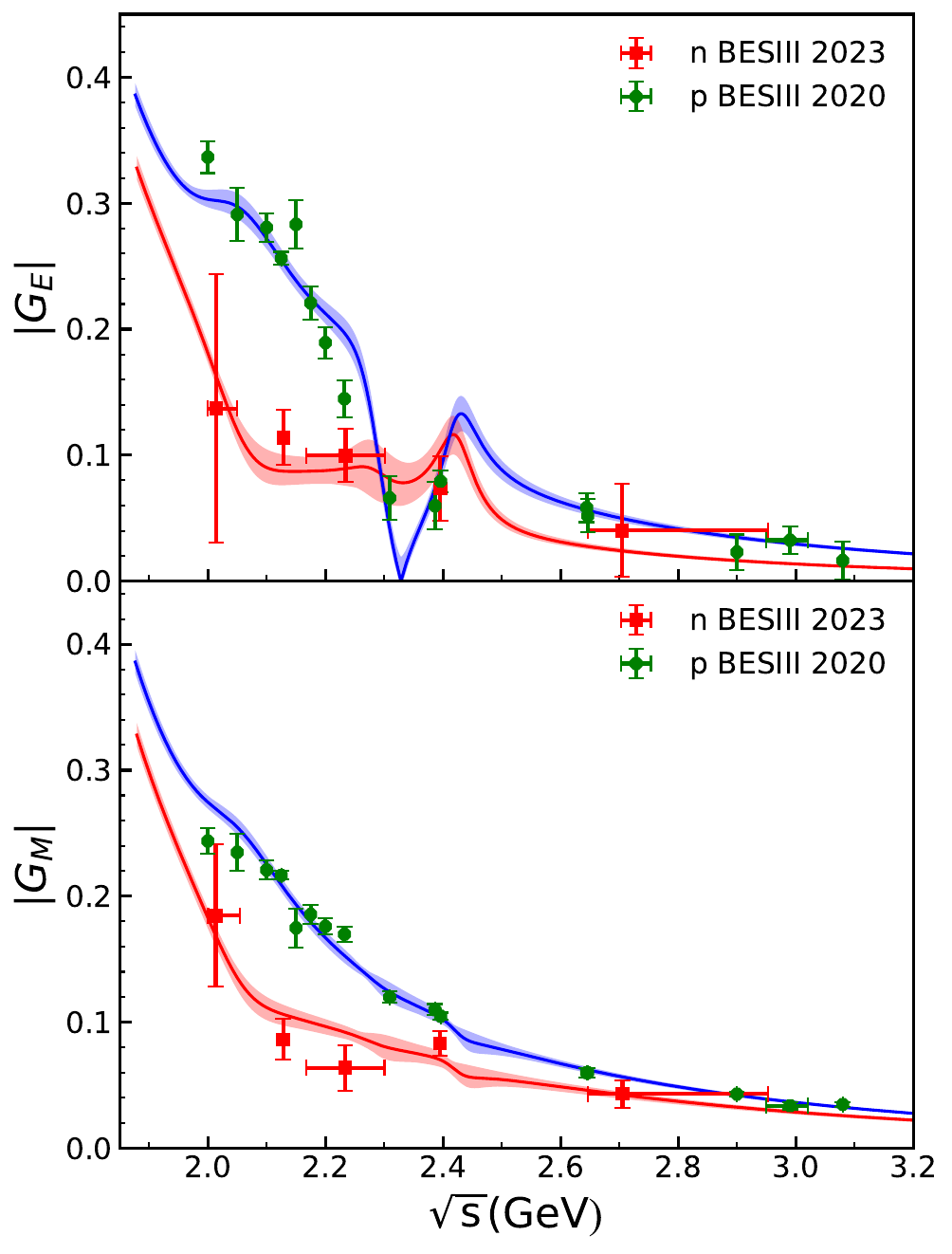}
	\caption{Fitted results for the electromagnetic form factors of the proton and neutron comparing with the experimental data, which are taken from: BESIII 2020~\cite{BESIII:2019hdp} and BESIII 2023~\cite{BESIII:2022rrg}. The error bands are calculated with the uncertainties of the fitted parameters.}   \label{fig:EMFFs}
\end{figure}

We show firstly the fitted numerical results of the electromagnetic form factors of proton and neutron in Fig.~\ref{fig:EMFFs}, comparing with the experimental data of the BESIII Collaboration. Within the contributions of the excited $\rho(2D)$, $\omega(3D)$, and $\omega(5S)$ states, we can obtain a good description of the electromagnetic form factors for both proton and neutron. For the electric form factor of proton, there is a dip structure around $2.3$ GeV which is because of the interference between the contributions of $\omega(3D)$ and $\omega(5S)$, and there is a clear peak for the $\omega(5S)$. This may indicate that, regarding to the current experimental data, the contribution of $\omega(5S)$ is necessary. It is expected that these theoretical results can be tested by future experiments.

\begin{figure}[htbp]
    \centering
    \includegraphics[scale=0.32]{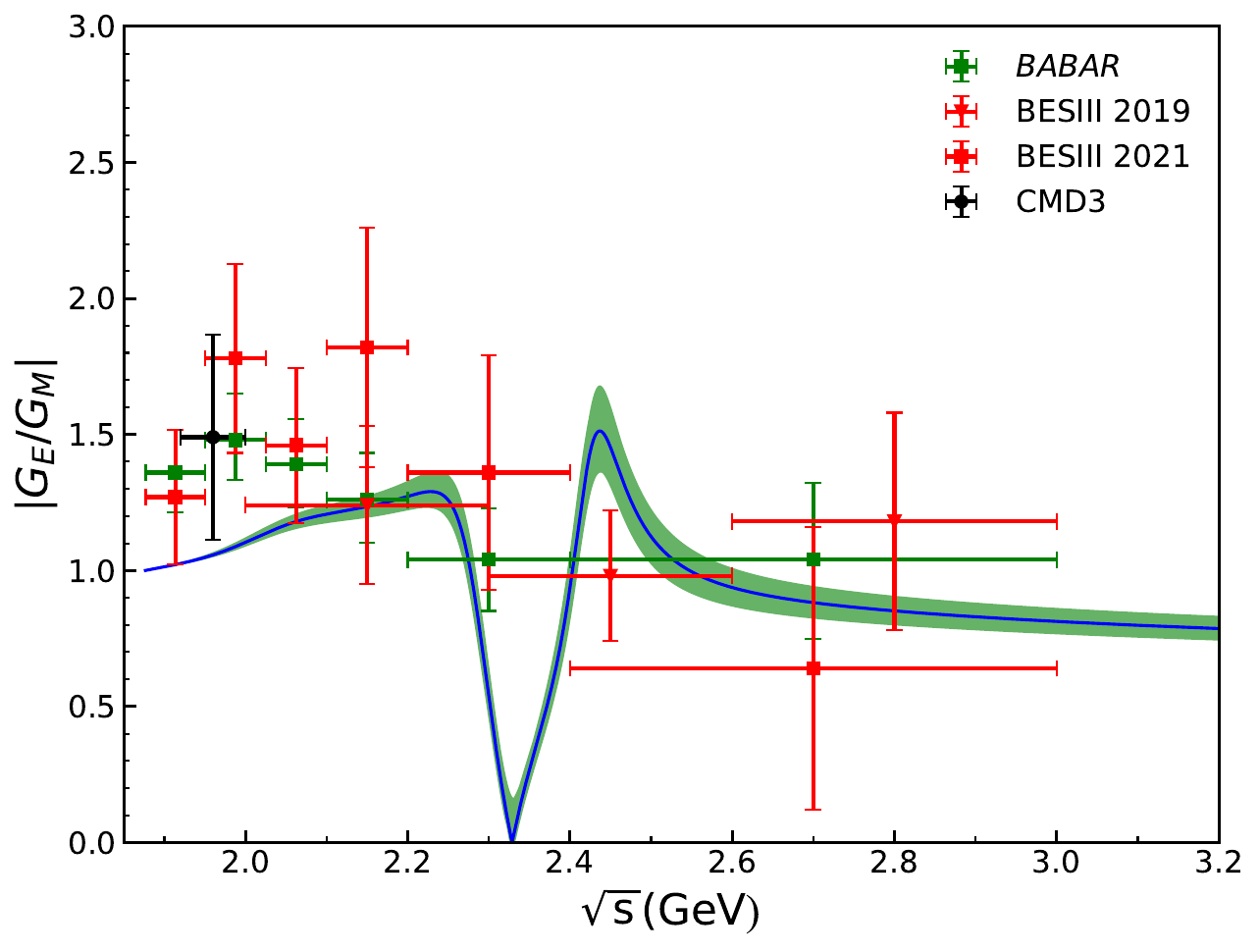}
    \caption{The ratio $|G_E/G_M|$ of proton comparing with the experimental data, which are taken from: $BABAR$~\cite{BaBar:2013ves}, BESIII 2019~\cite{BESIII:2019tgo}, BESIII 2021~\cite{BESIII:2021rqk}, CMD3~\cite{CMD-3:2015fvi}.}
    \label{fig:ratioGEGM}
\end{figure}

Second, we show the fitted results of the modulus of the ratio $G_E/G_M$ for the proton in Fig.~\ref{fig:ratioGEGM} compared with the experimental measurements. One can see that the experimental data have large errors. Again, there are dip and bump structures because of the contributions of of $\omega(3D)$ and $\omega(5S)$ states and their inteference. It is worthy mentioning that the ratio of $|G_E/G_M|$ equals to one at the reaction threshold as it should be. Clearly, more precise experimental data are needed to check our model calculations.

\begin{figure}[htbp]
	\centering
	\includegraphics[scale=0.4]{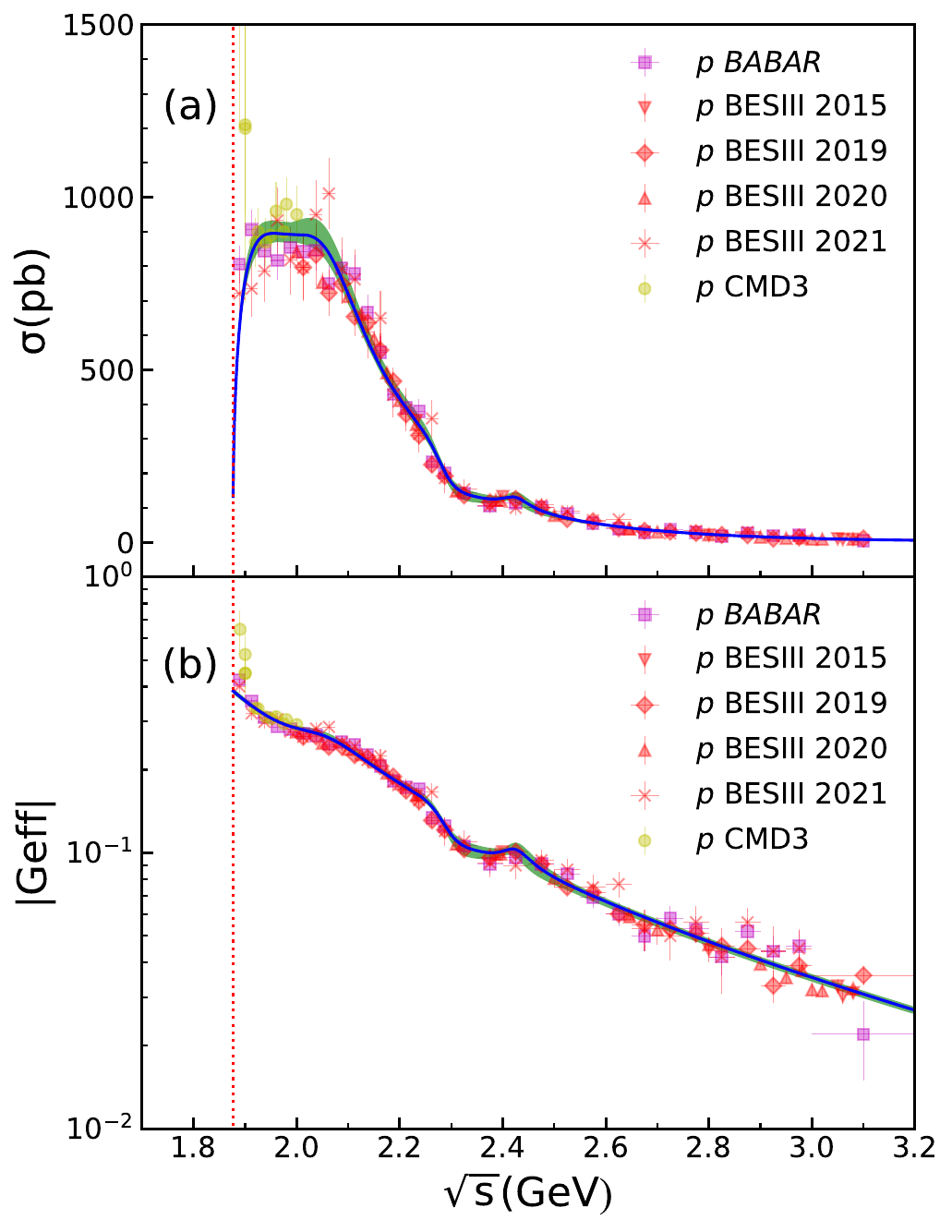}
	\caption{The fitted results of total cross sections (a) and the effective form factors (b) for proton. The theoretical error bands are obtained with the uncertainties of the fitted parameters. The red vertical dotted curve represent the $e^+ e^- \to p\bar{p}$ kinematic reaction threshold. The experimental data are taken from: $BABAR$~\cite{BaBar:2013ves}, BESIII 2015~\cite{BESIII:2015axk}, BESIII 2019~\cite{BESIII:2019tgo}, BESIII 2020~\cite{BESIII:2019hdp}, BESIII 2021~\cite{BESIII:2021rqk}, and CMD3~\cite{CMD-3:2015fvi}.}
	\label{fig:fittedresults-proton}
\end{figure}

\begin{figure}[htbp]
	\centering
	\includegraphics[scale=0.4]{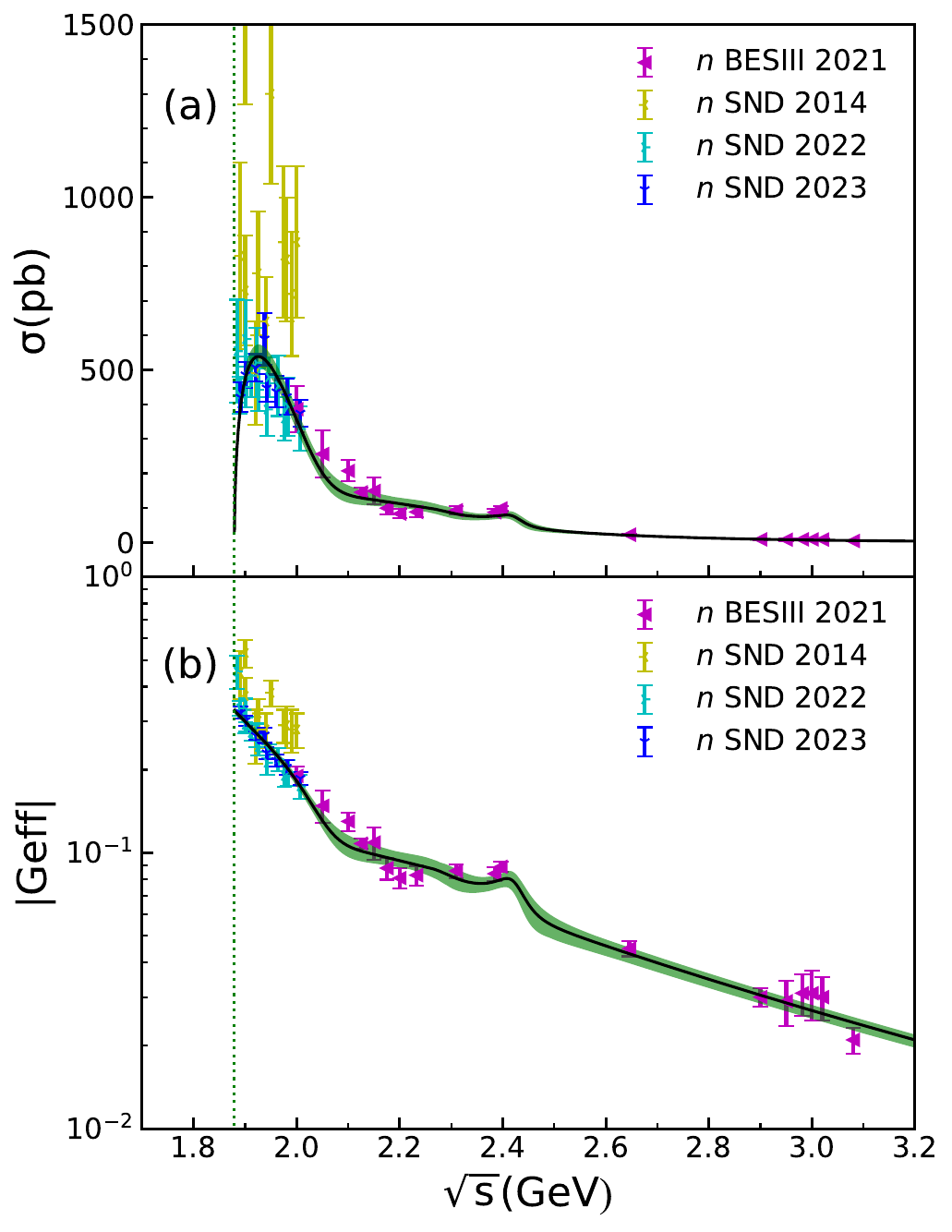}
	\caption{As in Fig.~\ref{fig:fittedresults-proton} but for the case of neutron. The experimental data are taken from: BESIII 2021~\cite{BESIII:2021tbq}, SND 2014~\cite{Achasov:2014ncd}, SND 2022~\cite{SND:2022wdb}, and SND 2023~\cite{SND:2023fos}.}
	\label{fig:fittedresults-neutron}
\end{figure}

The fitted numerical results of the total cross sections of $e^+e^-\to p \bar{p}$ and $e^+ e^- \to n \bar{n}$ reactions are shown in Fig.~\ref{fig:fittedresults-proton} (a) and Fig.~\ref{fig:fittedresults-neutron} (a), respectively. One can see that, thanks to the contribution of $\rho(2D)$, the platform behavior of the total cross section of $e^+e^-\to p \bar{p}$ can be well reproduced, and the experimental data of the $e^+ e^- \to n \bar{n}$ can also be described within the fitted model parameters. Furthermore, these nonmonotonic structures around $2.2$ to $2.4$ GeV can be fairly well reproduced by considering the contributions of $\omega(3D)$ and $\omega(5S)$ states. In fact, a resonance with mass around 2300 MeV and width about 188 MeV is needed in the analysis of the total cross sections of $e^+ e^- \to p\bar{p}$ reaction in Ref.~\cite{Xia:2021agf}, where a simple Breit-Wigner amplitude for the $e^+e^- \to p \bar{p}$ reaction was taken.

Since we can reproduce the total cross sections of $e^+e^-\to p \bar{p}$ and $e^+ e^- \to n \bar{n}$ reactions, it is expected that the effective form factors of proton and neutron can also be described well, as shown in Fig.~\ref{fig:fittedresults-proton} (b) and Fig.~\ref{fig:fittedresults-neutron} (b), respectively. This indicates that the nonmonotonic line shapes of the nucleon effective form factors can be explained within the VMD model, where the contributions of excited vector mesons are taken into account.

Before proceeding with further discussions, we need to clarify a few points. The excited vectors states of $\omega(3D)$ and $\omega(5S)$ can be replaced by the excited states of $\rho(3D)$ and $\rho(5S)$. This substitution does not significantly change the fitted results. As shown in Ref.~\cite{Wang:2021gle}, the calculated masses of $\rho(3D)$ and $\rho(5S)$ are degenerate with $\omega(3D)$ and $\omega(5S)$, respectively, while the widths of $\rho(3D)$ and $\rho(5S)$ are $158$ and $80$ MeV, respectively.

\begin{figure}[htbp]
	\centering
	\includegraphics[scale=0.4]{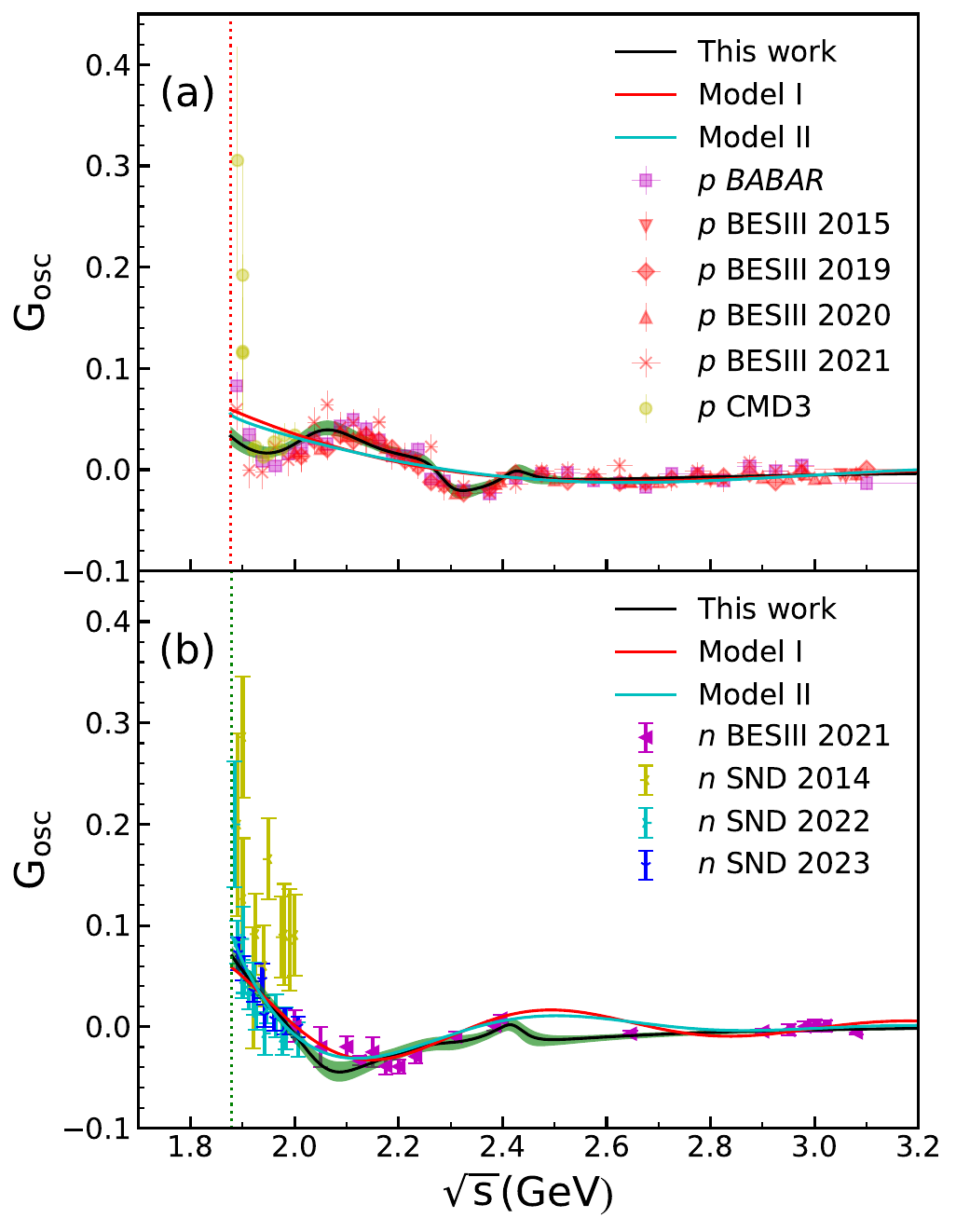}
	\caption{The fitted results of the oscillation part of the effective form factors for proton (a) and neutron (b). These error bands of "This work" are obtained with the uncertainties of the fitted parameters. The red and green vertical dotted curves represent the $e^+ e^- \to p\bar{p}$ and $n\bar{n}$ kinematic reaction threshold, respectively. The experimental data are taken from: $BABAR$~\cite{BaBar:2013ves}, BESIII 2015~\cite{BESIII:2015axk}, BESIII 2019~\cite{BESIII:2019tgo}, BESIII 2020~\cite{BESIII:2019hdp}, CMD3~\cite{CMD-3:2015fvi}, BESIII 2021~\cite{BESIII:2021rqk,BESIII:2021tbq}, SND 2014~\cite{Achasov:2014ncd}, SND 2022~\cite{SND:2022wdb}, and SND 2023~\cite{SND:2023fos}.}
	\label{fig:oscillation}
\end{figure}

Next, we will discuss the nonmonotonic features of the nucleon effective form factors. In general, the main part of the effective form factors of the nucleon can be described by a dipole form as shown in Eq.~\eqref{eq:dipoleform} with a global factor $c_0$, $G_D = c_0/(1-\gamma s)^2$. By fitting the experimental data of the effective form factors of proton and neutron, we get $c_0=5.54 \pm 0.02$ for proton and $c_0=4.08 \pm 0.04$ for neutron. Then, subtracting $G_D$ from the obtained effective form factor $|G_{\rm eff}|$, one can get the residual $G_{\rm osc}$,
\begin{eqnarray}
	G_{\rm osc} = |G_{\rm eff}| - G_{D} = |G_{\rm eff}| - \frac{c_0}{(1-\gamma s)^2}. \label{eq:osc}
\end{eqnarray}
The predictions of our model for the oscillation parts of the proton and neutron effective form factors are shown in Fig.~\ref{fig:oscillation} (a) and (b), respectively. It is found that the model predictions are in agreement with the subtracted data obtained with Eq.~\eqref{eq:osc}. One the other hand, for the oscillation part, we have performed new fitting with two formulas: $G_{\mathrm{osc}}=A \cos(C\sqrt{s}+D)/(1-\gamma s)^2$~\cite{Dai:2021yqr} and $G_{\rm{osc}} = A\exp (-Bp)\cos (Cp+D)$~\cite{BESIII:2021tbq} with $p$ the relative momentum between $N$ and $\bar{N}$ in the final state. The first one is formally referred to as Model I, while the other is designated as Model II. The fitted results are also shown in Fig.~\ref{fig:oscillation} (a) and (b). One can see that, with the dipole form of $G_{\rm{D}} $, the oscillation part of proton can not be well described, and the two fitting results are almost the same. While for the neutron, the oscillation part can be fairly well reproduced, and the two fittings are very similar. 

Note that, in Ref.~\cite{BESIII:2021tbq}, to get the conclusion which the effective form factor of both proton and neutron show a periodic behavior, different formula of $G_D$ were taken for proton and neutron. A three-pole formula $F_{3p} = F_0/[(1+s/s_0)(1-\gamma s)^2]$ with $s_0 = 8.8 \ \rm GeV^2$ were used for the case of proton~\cite{Tomasi-Gustafsson:2020vae,Tomasi-Gustafsson:2022tpu}, while the above dipole formula was used for neutron. In this work, we take the same formula for both proton and neutron. From the results shown in Fig.~\ref{fig:oscillation}, it is difficult to conclude the oscillation behaviour of the effective form factors of nucleon.~\footnote{Same result is obtained if we used the three-pole formula for both proton and neutron. The dipole form was firstly obtained from the best fit to the electromagnetic form factors of proton in the space-like region. Nevertheless, the VMD model and the dipole form of $g(q^2)$ can give a reasonable description of the experimental data on the baryon EMFFs at the considered energy region.}

Now we turn to the ratio of the total cross sections of $e^+ e^- \to p\bar{p}$ and $e^+ e^- \to n\bar{n}$ reactions. Until a few years ago, there had been some pioneering experimental efforts dedicated to the $e^+ e^- \to n\bar{n}$ reaction by the FENICE experiment~\cite{Antonelli:1998fv}. Even though the experimental data have large errors, and it is found that the total cross section of $e^+ e^- \to n\bar{n}$ reaction is twice as large as the one of $e^+ e^- \to p\bar{p}$ reaction, which is quite difficult to be understood, since most theoretical predictions are opposite. For example, the ratio of $R = \sigma_{e^+e^- \to p \bar{p}}/\sigma_{e^+ e^- \to n \bar{n}}$ is $7/3$ predicted by the SU(6) symmetric nucleon wave function~\cite{Farrar:1975yb} and it is the naive $e^2_u/e^2_d = 4$ from the quark charge ratio in the constituent quark model~\cite{Chernyak:1984bm}.

Very recently, the BESIII Collaboration has published a new precise measurement of the total cross sections of $e^+ e^- \to n\bar{n}$ reaction~\cite{BESIII:2021tbq}, and it is found that the ratio $R=\sigma_{e^+ e^- \to p \bar{p}}/\sigma_{e^+e^- \to n \bar{n}}$ is larger than 1 at all the measured energies, which is contrast to the results obtained by the FENICE experiment. The new results of BESIII Collaboration indicates that the photon-proton coupling is stronger than the corresponding photon-neutron coupling. The SND Collaboration also measured the total cross sections of $e^+ e^- \to n\bar{n}$ reaction for 14 energy points~\cite{SND:2022wdb} and their experimental results are consistent with the results of BESIII Collaboration.

\begin{figure}[htbp]
    \centering
    \includegraphics[scale=0.35]{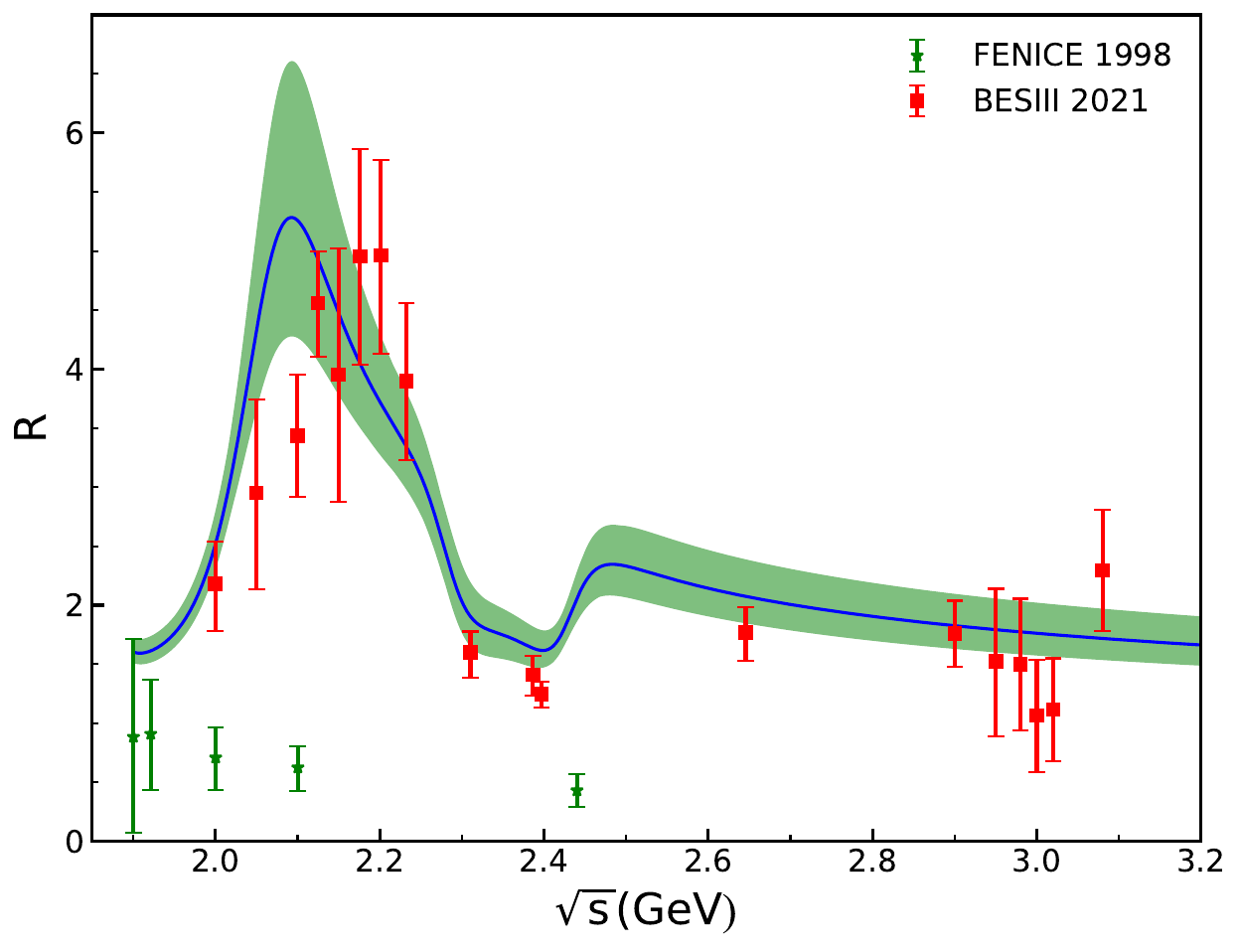}
    \caption{The ratio $R = \sigma_{e^+ e^- \to p \bar{p}}/\sigma_{e^+e^- \to n \bar{n}}$ as a function of $\sqrt{s}$ compared with the experimental data which are taken from: FENICE 1998~\cite{Antonelli:1998fv} and BESIII 2021~\cite{BESIII:2021tbq}.}  \label{fig:TCSratio}
\end{figure}

Our theoretical results of the ratio $R$ are shown in Fig.~\ref{fig:TCSratio} by blue curve with error bands obtained from the uncertainties of the fitted parameters. The experimental data taken from BESIII and FENICE are also shown. Again, we can describe quite well the new experimental data by BESIII Collaboration~\cite{BESIII:2021tbq}.

\begin{figure}[htbp]
	\centering
	\includegraphics[scale=0.35]{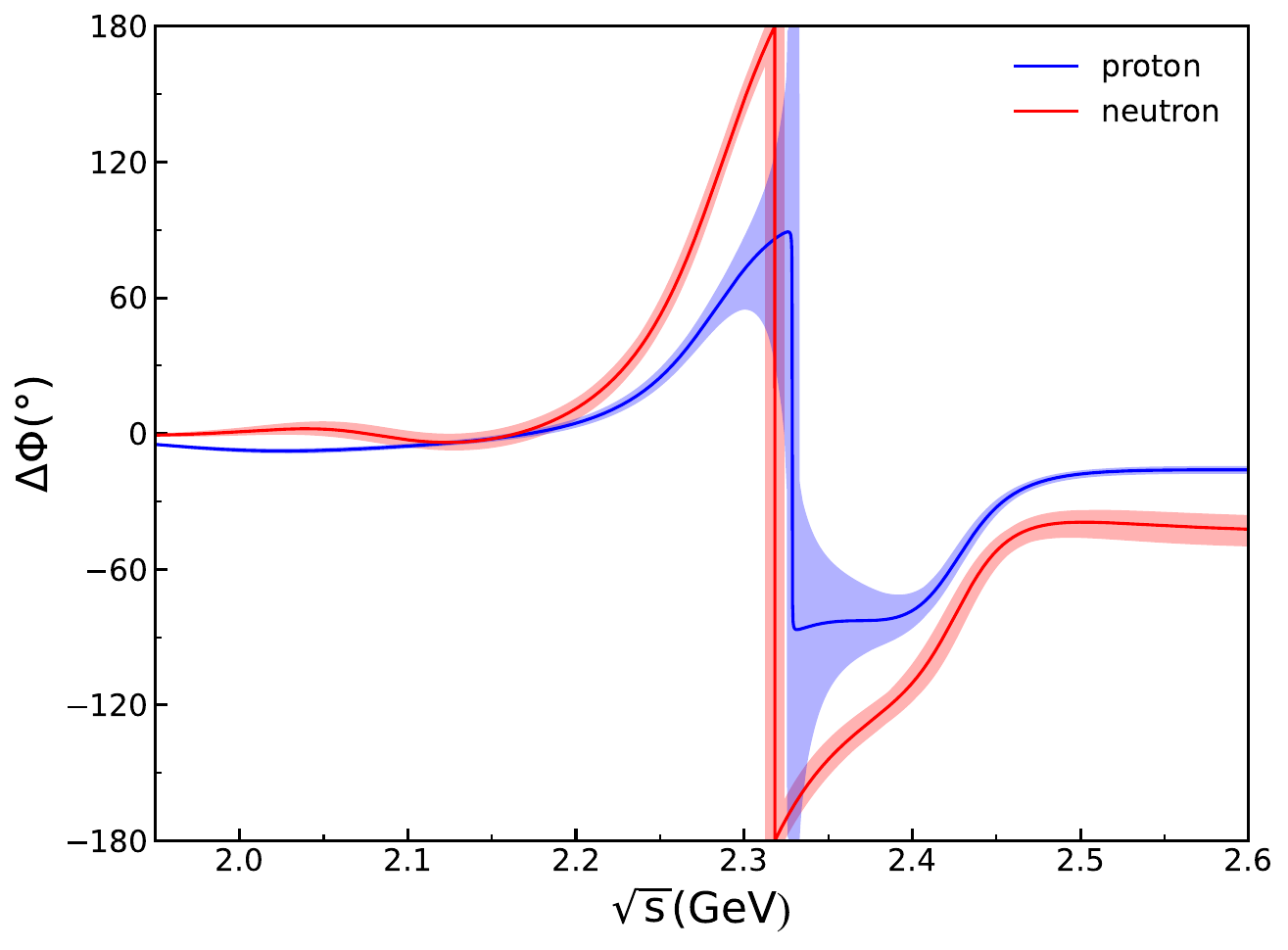}
	\caption{The predictions of $\Delta \Phi$ for proton and neutron as a function of $\sqrt{s}$. The error bands are obtained with the uncertainties of fitted parameters.}  \label{fig:deltaphi}
\end{figure}

Moreover, we also calculated the relative phase $\Delta \Phi$ between $G_E$ and $G_M$ of the nucleon. The relative phase $\Delta\Phi$ is related to the spin polarization of baryons in the $e^+e^- \to B\bar{B}$ reaction~\cite{Faldt:2017kgy,BESIII:2020fqg,BESIII:2021cvv}. Here, we can write

\begin{align}
	G_E/G_M  = e^{i \Delta \Phi} |G_E/G_M|.
\end{align}
Using the fitted parameters in Table~\ref{tab:fittedparameters}, we can obtain the $\Delta \Phi$ as a function of $\sqrt{s}$ for both proton and neutron, which are shown in Fig.~\ref{fig:deltaphi}. In the reaction threshold, the relative phase $\Delta \Phi$ equals 0, since the electric and magnetic form factors are equivalent. As the center of mass energy $\sqrt{s}$  increases, the value of $\Delta \Phi$ undergoes significant changes, particularly in the vicinity of $\sqrt{s} = 2.30 $ GeV. This is because the value of $G_E$ for proton is close to zero and it is very small for the neutron. Since the strength of the polarization is proportional to the $\sin \Delta \Phi$, according to the numerical results as shown in Fig.~\ref{fig:deltaphi}, it implies that there is obvious spin polarization at 2.28 GeV and 2.40 GeV, and the spin polarization has opposite directions at the two energy points. It is expected that the future experiments can be used to check our model calculations.

\section{SUMMARY}

In this work, we study the effective form factors of proton and neutron in the timelike region using a phenomenological extending vector meson dominance model. In addition to the contribution of the ground states $\rho$ and $\omega$, the contributions from three theoretical predicted vector states $\rho(2D)$, $\omega(3D)$, and $\omega(5S)$ are also taken into account. It is found that we can successfully describe the cross sections of $e^+ e^- \to p \bar{p}$ and $e^+ e^- \to n \bar{n}$ reactions and also the effective form factors for both proton and neutron.  Furthermore, the electromagnetic form factors and the modulus of their ratio can be also obtained, which are in agreement with the experimental data. In addition, thanks to the contributions from the excited vector resonances and the phenomenological dipole formula for the intrinsic form factor $g(s) = 1/(1-\gamma s)^2$, the flat behaviour of the nucleon effective form factors around the threshold can be reproduced, and we are able to provide an natural explanation for the nonmonotonic behavior which was found in the time-like nucleon electromagnetic form factors.

Our study here shows that the so-called periodic behaviour of the nucleon effective form factors is not confirmed. However, there are indeed nonmonotonic structures in the line shape of nucleon effective form factors, which can be naturally reproduced by considering the contributions from the low-lying excited vector states. We hope that our present work can stimulate more studies along this line, and it is expected that more precise experimental measurements in near future, especially for the case of the $e^+ e^- \to n\bar{n}$ reaction, can be used to check our model calculations.

Finally, we would like to stress that, thanks to the
important role played by the excited vector states contribution in the
$e^+ e^- \to N\bar{N}$ reaction, accurate data for this reaction
can be used to improve our knowledge of those low-lying excited vector states above 2 GeV, which are at present poorly known. This work
constitutes a first step in this direction.

\section*{ACKNOWLEDGEMENTS}

We would like to thank Ji-Feng Hu and Lei Xia for useful discussions. This work is partly supported by the National Natural Science Foundation of China under Grant No. 12075288. It is also supported by the Youth Innovation Promotion Association CAS.

\bibliography{ref.bib}

\end{document}